\documentclass[submission, Proceedings]{SciPost}

\hypersetup{
    colorlinks,
    linkcolor={red!50!black},
    citecolor={blue!50!black},
    urlcolor={blue!80!black}
}

\usepackage[bitstream-charter]{mathdesign}
\usepackage{mathtools}
\usepackage[capitalise]{cleveref}  
\usepackage{braket}
\usepackage[percent]{overpic}  
\usepackage{wasysym}  
\definecolor{GreenNoam}{rgb}{0,0.5,0}
\usepackage[colorinlistoftodos, textsize=small, 
backgroundcolor=blue!20!white, bordercolor=red, 
linecolor=red]{todonotes}

\DeclareSymbolFont{usualmathcal}{OMS}{cmsy}{m}{n}
\DeclareSymbolFontAlphabet{\mathcal}{usualmathcal}

\begin{document}

\begin{center}{\Large \textbf{
An algebraic approach to intertwined quantum
phase transitions in the Zr isotopes\\
}}\end{center}

\begin{center}
N. Gavrielov\textsuperscript{1,2$\star$}
\end{center}

\begin{center}
{\bf 1} Center for Theoretical Physics, Sloane Physics 
Laboratory, Yale University, New Haven, Connecticut 
06520-8120, USA
\\
{\bf 2} Racah Institute of Physics, The Hebrew 
University, Jerusalem 91904, Israel
\\

* noam.gavrielov@yale.edu
\end{center}

\begin{center}
\today
\end{center}


\definecolor{palegray}{gray}{0.95}
\begin{center}
\colorbox{palegray}{
  \begin{tabular}{rr}
  \begin{minipage}{0.1\textwidth}
    \includegraphics[width=20mm]{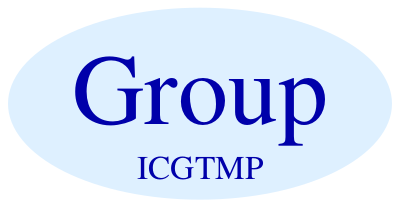}
  \end{minipage}
  &
  \begin{minipage}{0.85\textwidth}
    \begin{center}
    {\it 34th International Colloquium on Group Theoretical Methods in Physics}\\
    {\it Strasbourg, 18-22 July 2022} \\
    \doi{10.21468/SciPostPhysProc.?}\\
    \end{center}
  \end{minipage}
\end{tabular}
}
\end{center}

\section*{Abstract}
{\bf
The algebraic framework of the interacting boson model with 
configuration mixing is employed to demonstrate the 
occurrence of intertwined quantum phase transitions (IQPTs) 
in the $_{40}$Zr isotopes with neutron number 52--70. The 
detailed quantum and classical analyses reveal a QPT of 
crossing normal and intruder configurations superimposed on 
a QPT of the intruder configuration from U(5) to SU(3) and 
a crossover from SU(3) to SO(6) dynamical symmetries.
}

\vspace{10pt}
\noindent\rule{\textwidth}{1pt}
\tableofcontents\thispagestyle{fancy}
\noindent\rule{\textwidth}{1pt}
\vspace{10pt}

\section{Introduction}\label{sec:intro}
Quantum phase transitions 
\cite{Gilmore1978a,Gilmore1979,carr2010QPT} 
are qualitative changes in the structure of a physical 
system that occur as a function of one (or more) parameters 
that appear in the quantum Hamiltonian describing the 
system.
In nuclear physics~\cite{Cejnar2010}, we vary the number of 
nucleons and examine mainly two types of quantum phase 
transitions (QPTs).
The first describes shape phase transitions in a single 
configuration, denoted as Type I. When interpolating 
between two shapes, for example, the Hamiltonian can be 
written as a sum of two parts
\begin{equation}\label{eq:type-i}
\hat H = (1-\xi)\hat H_1 + \xi\hat H_2~,
\end{equation}
with $\xi$ the control parameter. As we vary $\xi$ with 
nucleon 
number from 0 to 1, the equilibrium shape and symmetry of 
the 
Hamiltonian vary from those of $\hat H_1$ to those of $\hat 
H_2$. QPTs of this type have been studied extensively in 
the 
framework of the interacting boson model (IBM) 
\cite{Dieperink1980,Cejnar2009,Cejnar2010,Iachello2011}.
One example of such QPT is the $_{62}$Sm region with 
neutron number 84--94, where the shape evolves from 
spherical to axially-deformed, with a critical point at 
neutron number 90.

The second type of QPT occurs when the ground state 
configuration changes its character, typically from normal 
to 
intruder type of states, denoted as Type~II QPT. In such 
cases, 
the Hamiltonian can be written in matrix form 
\cite{Frank2006}. 
For two configurations $A$ and $B$ we have
\begin{equation}\label{eq:type-ii}
\hat H =
\left [
\begin{array}{cc}
\hat H_A(\xi_A) & \hat W(\omega) \\ 
\hat W(\omega)  & \hat H_B(\xi_B)
\end{array}
\right]~,
\end{equation}
with $\xi_i$ ($i\!=\!A,B$), the control parameter of 
configuration ($i$), and $\hat W$, the coupling between 
them with parameter $\omega$. QPTs of this type are 
manifested empirically near (sub-) shell closure, e.g. in 
the light Pb-Hg isotopes, with strong mixing between the 
configurations~\cite{GarciaRamos2009,GarciaRamos2015a}.

Recently, we have introduced a new type of 
phase-transitions in even-even 
\cite{Gavrielov2019,Gavrielov2022} and odd-mass 
\cite{Gavrielov2022c} nuclei called intertwined quantum 
phase 
transitions (IQPTs). The latter refers to a scenario where 
as we vary the control parameters ($\xi_A, \xi_B, \omega$) 
in \cref{eq:type-ii}, each of the Hamiltonians $\hat H_A$ 
and $\hat H_B$ undergoes a separate and clearly 
distinguished shape-phase transition (Type~I), and the 
combined Hamiltonian simultaneously experiences a crossing 
of configurations $A$ and $B$ (Type~II).

\section{Theoretical framework}\label{sec:theo-frame}
A convenient framework to study the different types of QPTs 
together is the extension of the IBM to include 
configuration 
mixing 
(IBM-CM)~\cite{IachelloArimaBook,Duval1981,Duval1982}.

\subsection{The interacting boson model with configuration 
mixing}\label{sec:ibm}
The IBM for a single shell model configuration has been 
widely used to describe low-lying quadrupole collective 
states in nuclei in terms of $N$ monopole ($s^\dagger$) and 
quadrupole ($d^\dagger$) bosons, representing valence 
nucleon pairs. 
The model has U(6) as a spectrum generating algebra, where 
the Hamiltonian is expanded in terms of its generators, 
$\{s^\dagger s, s^\dagger d_\mu, d^\dagger_\mu s, 
d^\dagger_\mu d_{\mu^\prime}\}$, and consists of Hermitian, 
rotational-scalar interactions which conserve the total 
number of $s$- and $d$-bosons $\hat N=\hat n_s+\hat 
n_d\!=\allowbreak\!s^\dagger s+\sum_\mu d^\dagger_\mu 
d_\mu~$. 
The boson number is fixed by the microscopic interpretation 
of the IBM~\cite{IachelloTalmi1987} to be 
$N\!=\!N_{\pi}+N_{\nu}$, where $N_{\pi}$ ($N_{\nu}$) is the 
number of proton (neutron) particle or hole pairs counted 
from the nearest closed shell. 

The solvable limits of the model correspond to dynamical 
symmetries (DSs) associated with chains of nested 
sub-algebras of U(6), terminating in the invariant SO(3) 
algebra. In the IBM there are three DS limits
\begin{align}\label{eq:ds-chains}
\text{U(6)} \supset \begin{cases}
	  &\text{U(5)} \supset \text{SO(5)} \supset 
	  \text{SO(3)},\\
      &\text{SU(3)} \supset \text{SO(3)}, \\
	  &\text{SO(6)} \supset \text{SO(5)} \supset 
	  \text{SO(3)}.
\end{cases}
\end{align}
In a DS, the Hamiltonian is written in terms of Casimir 
operators of the algebras of a given chain. In such a case, 
the 
spectrum is completely solvable and resembles known 
paradigms of 
collective motion: spherical vibrator [U(5)], axially 
symmetric 
[SU(3)] and $\gamma$-soft deformed rotor [SO(6)]. 
In each case, the energies and eigenstates are labeled by 
quantum numbers that are the labels of irreducible 
representations (irreps) of the algebras in the chain. The 
corresponding basis states for each of the chains 
\eqref{eq:ds-chains} are
\begin{subequations}
\begin{align}
\text{U(5)}: & \quad\ket{N,n_d,\tau,n_\Delta,L},\\
\text{SU(3)}: & \quad\ket{N,(\lambda,\mu),K,L},\\
\text{SO(6)}: & \quad\ket{N,\sigma,\tau,n_\Delta,L},
\end{align}
\end{subequations}
where $N,n_d,(\lambda,\mu),\sigma,\tau,L$ label the irreps 
of 
U(6), U(5), SU(3), SO(6), SO(5) and SO(3), respectively, 
and 
$n_\Delta,K$ are multiplicity labels.

An extension of the IBM to include intruder excitations is 
based on associating the different shell-model spaces of 
0p-0h, 2p-2h, 4p-4h, $\dots$ particle-hole excitations, 
with the corresponding boson spaces with $N,\, N\!+\!2,\, 
N\!+\!4,\ldots$ bosons, which are subsequently 
mixed~\cite{Duval1981,Duval1982}. For two configurations 
the resulting IBM-CM Hamiltonian can be transcribed in a 
form equivalent to that of \cref{eq:type-ii}
\begin{equation}\label{eq:ham-cm}
\hat H = \hat H_A^{(N)} + \hat H_B^{(N+2)} + \hat 
W^{(N,N+2)}~.
\end{equation}
Here, the notations $\hat{\cal 
O}^{(N)}\!=\!\hat{P}_{N}^{\dag}\hat{\cal O}\hat{P}_{N}$ and 
$\hat{\cal O}^{(N,N^{\prime})}\!=\!\hat{P}_{N}^{\dag 
}\hat{\cal O}\hat{P}_{N^{\prime }}$, stand for an operator 
$\hat{\cal O}$, with $\hat{P}_{N}$, a projection operator 
onto the $N$ boson space. The Hamiltonian 
$\hat{H}_{A}^{(N)}$ represents the $N$ boson space (normal 
$A$ configuration) and $\hat{H}_{B}^{(N+2)}$ represents the 
$N\!+\!2$ boson space (intruder $B$ configuration).

\subsection{Wave functions structure}\label{sec:conf-sym}
The eigenstates $\ket{\Psi;L}$ of the Hamiltonian 
\eqref{eq:ham-cm} with angular momentum $L$, are linear 
combinations of the wave functions, $\Psi_A$ and $\Psi_B$, 
in the two spaces $[N]$ and $[N+2]$,
\begin{equation}\label{eq:wf}
\ket{\Psi; L} = a\ket{\Psi_A; [N], L} + b\ket{\Psi_B; 
[N\!+\!2], L}\, ,\;
\end{equation}
with $a^{2} + b^{2} \!=\! 1$. We note that each of the 
components in \cref{eq:wf}, $\ket{\Psi_A; [N], L}$ and 
$\ket{\Psi_B; [N\!+\!2], L}$, can be expanded in terms of 
the different DS limits with its corresponding boson number 
in the following manner
\begin{equation}\label{eq:wf-ds}
\ket{\Psi_i; [N_i],L} = \sum_{\alpha} 
C^{(N_i,L)}_{\alpha}\ket{N_i,\alpha,L},
\end{equation}
where $N_A\!=\!N$ and $N_B\!=\!N+2$, and $\alpha= 
\{n_d,\tau,n_\Delta\}, \{(\lambda,\mu),K\}, 
\{\sigma,\tau,n_{\Delta}\}$ are the quantum numbers of the 
DS eigenstates. The coefficients $C^{(N,L)}_\alpha$ give 
the weight of each component in the wave function.
Using them, we can calculate the wave function probability 
of having definite quantum numbers of a given symmetry in 
the DS bases, \cref{eq:wf-ds}, for its 
$A$ or $B$ parts
\begin{subequations}\label{eq:decomp-ds}
\begin{align}
\text{U(5)}: \quad & P^{(N_i,L)}_{n_d} = 
\sum_{\tau,n_\Delta}[C^{(N_i,L)}_{n_d,\tau,n_\Delta}]^2, 
&\text{SO(6)}: \quad & P^{(N_i,L)}_{\sigma} = 
\sum_{\tau,n_\Delta}[C^{(N_i,L)}_{\sigma,\tau,n_\Delta}]^2,
\\
\text{SU(3)}: \quad & P^{(N_i,L)}_{(\lambda,\mu)} = 
\sum_{K}[C^{(N_i,L)}_{(\lambda,\mu),K}]^2,
&\text{SO(5)}: \quad & P^{(N_i,L)}_{\tau} = 
\sum_{n_d,n_\Delta}[C^{(N_i,L)}_{n_d,\tau,n_\Delta}]^2\label{eq:decomp-so5}.
\end{align}
\end{subequations}
Here the subscripts $i\!=\!A,B$ denote the different 
configurations, i.e., $N_A\!=\!N$ and $N_B\!=\!N+2$.
Furthermore, for each eigenstate \eqref{eq:wf}, we can 
also examine its coefficients $a$ and $b$, which portray 
the probability of the normal-intruder mixing. They are 
evaluated from the sum of the squared coefficients of an 
IBM basis. For the U(5) basis, we have
\begin{equation}\label{eq:decomp-cm}
P^{(N_A,L)}_a \equiv a^2 = \sum_{n_d,\tau,n_\Delta} 
|C^{(N_A,L)}_{n_d,\tau,n_\Delta}|^2; \qquad
P^{(N_B,L)}_b \equiv 
b^2 = \sum_{n_d,\tau,n_\Delta} 
|C^{(N_B,L)}_{n_d,\tau,n_\Delta}|^2.
\end{equation}
where the sum goes over all possible values of 
$(n_d,\tau,n_\Delta)$ in the $(N_i,L)$ space, $i=A,B$, and 
$a^2 + b^2\!=\!1$.

\subsection{Geometry}\label{sec:geometry}
To obtain a geometric interpretation of the IBM is we take 
the
expectation value of the Hamiltonian between  coherent 
(intrinsic) states~\cite{Ginocchio1980a,Dieperink1980} 
to form an energy surface
\begin{align}\label{eq:surface}
E_N(\beta,\gamma) &=
\bra{\beta,\gamma; N} \hat H \ket{\beta,\gamma;N}~.
\end{align}
The $(\beta,\gamma)$ of \cref{eq:surface} are quadrupole 
shape 
parameters whose values, $(\beta_{\rm eq},\gamma_{\rm 
eq})$, at 
the global minimum of $E_{N}(\beta,\gamma)$ define the 
equilibrium shape for a given Hamiltonian. The values are 
$(\beta_{\rm eq}\!=\!0)$, $(\beta_{\rm 
eq}\!=\!\sqrt{2},\gamma_{\rm eq}\!=\!0)$ and $(\beta_{\rm 
eq}\!=\!1,\gamma_{\rm eq}\text{~arbitrary})$ for the U(5), 
SU(3) 
and SO(6) DS limits, respectively. Furthermore, for these 
values 
the ground-band intrinsic state, $\ket{\beta_{\rm 
eq},\gamma_{\rm eq};N}$, becomes a lowest weight state in 
the irrep of the leading subalgebra of the 
DS chain, with quantum numbers $(n_d\!=\!0)$, 
$(\lambda,\mu)\!=\!(2N,0)$ and $(\sigma\!=\!N)$ for the 
U(5), SU(3) and SO(6) DS limits, respectively.

For the IBM-CM Hamiltonian, the energy surface takes a 
matrix form \cite{Frank2004}
\begin{align}\label{eq:surface-mat}
E(\beta,\gamma) =
\left [
\begin{array}{cc}
E_A(\beta,\gamma;\xi_A) & \Omega(\beta,\gamma;\omega) \\ 
\Omega(\beta,\gamma;\omega) & E_B(\beta,\gamma;\xi_B)
\end{array}
\right ] ,
\end{align} 
where the entries are the matrix elements of the 
corresponding terms in the Hamiltonian~\eqref{eq:type-ii}, 
between the intrinsic states of each of the configurations, 
with the appropriate boson number. Diagonalization of 
this two-by-two matrix produces the so-called 
eigen-potentials, $E_{\pm}(\beta,\gamma)$.

\subsection{QPTs and order 
parameters}\label{sec:order-param}
The energy surface depends also on the Hamiltonian 
parameters and serves as the Landau potential whose 
topology determines the type of phase transition.
In QPTs involving a single configuration (Type I), the 
ground state shape defines the phase of the system, which 
also identifies the corresponding DS as the phase of the 
system. 
Such Type~I QPTs can be studied using a Hamiltonian 
as in \cref{eq:type-i}, that interpolates between different 
DS 
limits (phases) by varying its control parameters $\xi$. 
The order parameter is taken to be the expectation value of 
the $d$-boson number operator, $\hat n_d$, in the ground 
state, $\braket{\hat n_d}_{0^+_1}$, and measures the amount 
of deformation in the ground state.

In QPTs involving multiple configurations (Type II), the 
dominant configuration in the ground state defines the 
phase of the system. Such Type~II QPTs can be studied using 
a 
Hamiltonian as in \cref{eq:ham-cm}, that interpolates 
between 
the different configurations by varying its control 
parameters 
$\xi_A,\xi_B,\omega$. The order parameters are taken to be 
the expectation value of $\hat n_d$ in the ground state 
wave function, $\ket{\Psi;L=0^+_1}$, and in its $\Psi_A$ 
and $\Psi_B$ components, \cref{eq:wf}, denoted by 
$\braket{\hat n_d}_{0^+_1}$, $\braket{\hat n_d}_A$ and 
$\braket{\hat n_d}_B$, respectively. The shape-evolution in 
each of the configurations~$A$ and $B$ is encapsulated in 
$\braket{\hat n_d}_A$ and $\braket{\hat n_d}_B$, 
respectively. Their sum weighted by the probabilities of 
the $\Psi_A$ and $\Psi_B$ components $\braket{\hat 
n_d}_{0^+_1} = a^2\braket{\hat n_d}_A + b^2\braket{\hat 
n_d}_B$, portrays the evolution of the normal-intruder 
mixing.

\section{QPTs in the Zr isotopes}\label{sec:qpt-zr}
Along the years, the $Z\approx40,~A\approx100$ region was 
suggested by many works to have a ground state that is 
dominated by a normal spherical configuration for neutron 
numbers 50--58 and by an intruder deformed configuration 
for 60 
onward.
This dramatic change in structure is explained in the shell 
model by the isoscalar proton-neutron interaction between 
non-identical nucleons that occupy the spin-orbit 
partner orbitals $\pi 1g_{9/2}$ and $\nu 
1g_{7/2}$~\cite{Federman1979}. The crossing between 
configurations arises from the promotion of protons across 
the Z=40 subsell gap. The interaction energy results in a 
gain 
that compensates the loss in single-particle and pairing 
energy 
and a mutual polarization effect is enabled. Therefore, the 
single-particle orbitals at higher intruder configurations 
are 
lowered near the ground state normal configuration, which 
effectively reverses their order.
\subsection{Model space}
Using the framework of the IBM-CM, we consider 
$_{40}^{90}$Zr as 
a core and valence neutrons in the 50--82 major shell. The 
normal $A$ configuration corresponds to having no active 
protons 
above $Z\!=\!40$ sub-shell gap, and the intruder $B$ 
configuration corresponds to two-proton excitation from 
below to above this gap, creating 2p-2h states. 
Therefore, the IBM-CM model space employed in this study, 
consists of $[N]\oplus[N+2]$ boson spaces with total 
boson number $N=1,2,\ldots8$ for $^{92-106}$Zr and $\bar 
N=\bar 7,\bar 6$ for $^{108,110}$Zr, respectively, where 
the bar over a number indicates that these are hole bosons.
\subsection{Hamiltonian and $E2$ transitions operator}
In order to describe the spectrum of the Zr isotopes, we 
take a Hamiltonian that has a form as in \cref{eq:ham-cm} 
with entries
\begin{subequations}
\label{eq:ham_ab}
\begin{align}
\hat H_A(\epsilon^{(A)}_d,\kappa^{(A)},\chi)  & = 
\epsilon^{(A)}_d\, \hat n_d + \kappa^{(A)}\, \hat Q_\chi 
\cdot \hat Q_\chi~,
\label{eq:ham_a}
\\
\hat H_B(\epsilon^{(B)}_d,\kappa^{(B)},\chi) & =
\epsilon^{(B)}_d\, \hat n_d + \kappa^{(B)}\, \hat Q_\chi 
\cdot \hat Q_\chi + \kappa^{\prime(B)} \hat L \cdot \hat L 
+ \Delta_p~,
\label{eq:ham_b}
\end{align}
\end{subequations}
where the quadrupole operator is given by
$\hat Q_\chi =
d^\dag s+s^\dag \tilde d \!+\!\chi (d^\dag \times \tilde 
d)^{(2)}$, and $\hat L = 
\sqrt{10}(d^\dagger\tilde d)^{(1)}$ is the angular momentum 
operator. Here $\tilde d_m = (-1)^m d_{-m}$ and standard 
notation of angular momentum coupling is used. The off-set 
energy between configurations~$A$ and $B$ is $\Delta_p$, 
where 
the index $p$ denotes the fact that this is a proton 
excitation. 
The mixing term in 
\cref{eq:ham-cm} between configurations $(A)$ and $(B)$ has 
the form~\cite{IachelloArimaBook, Duval1981, Duval1982}
$\hat W = \omega\,[\,(d^\dag\times d^\dag)^{(0)} + 
(s^\dag)^2\,] + {\rm H.c.}$,
where H.c. stands for Hermitian conjugate. The parameters 
are 
obtained from a fit, elaborated in the appendix of 
Ref.~\cite{Gavrielov2022}.

The $E2$ operator for two configurations is written as
$\hat{T}(E2)\!=\!e^{(A)}\hat 
Q^{(N)}_{\chi}+e^{(B)}\hat Q^{(N+2)}_{\chi}$, with 
$\hat{Q}_{\chi}^{(N)}\!=\!\hat{P}_{N}^{\dag}
\hat{Q}_{\chi}\hat{P}_{N}$ and 
$\hat{Q}_{\chi}^{(N+2)}\!=\!P_{N+2}^{\dag}
\hat{Q}_{\chi}\hat{P}_{N+2}$. The boson effective charges 
$e^{(A)}$ and $e^{(B)}$ are determined from the 
$2^{+}\!\to\! 0^{+}$ transition within each configuration 
\cite{Gavrielov2022}, 
and $\chi$ is the same parameter as in the 
Hamiltonian~\eqref{eq:ham_ab}.

For the energy surface matrix \eqref{eq:surface-mat}, we 
calculate the expectation values of the Hamiltonians
$\hat H_A$~\eqref{eq:ham_a} and $\hat 
H_B$~\eqref{eq:ham_b} in the intrinsic 
state of \cref{sec:geometry} with $N$ and $N\!+\!2$ bosons 
respectively, and a non-diagonal matrix element of the 
mixing 
term $\hat W$ between them. The explicit expressions can be 
found in \cite{Gavrielov2022}.

\section{Results}
In order to understand the change in structure of the
Zr isotopes, it is insightful to examine the evolution of
different properties along the chain.
\subsection{Evolution of energy levels}
\begin{figure}[t!]
\centering
\begin{overpic}[height=0.2\textheight]{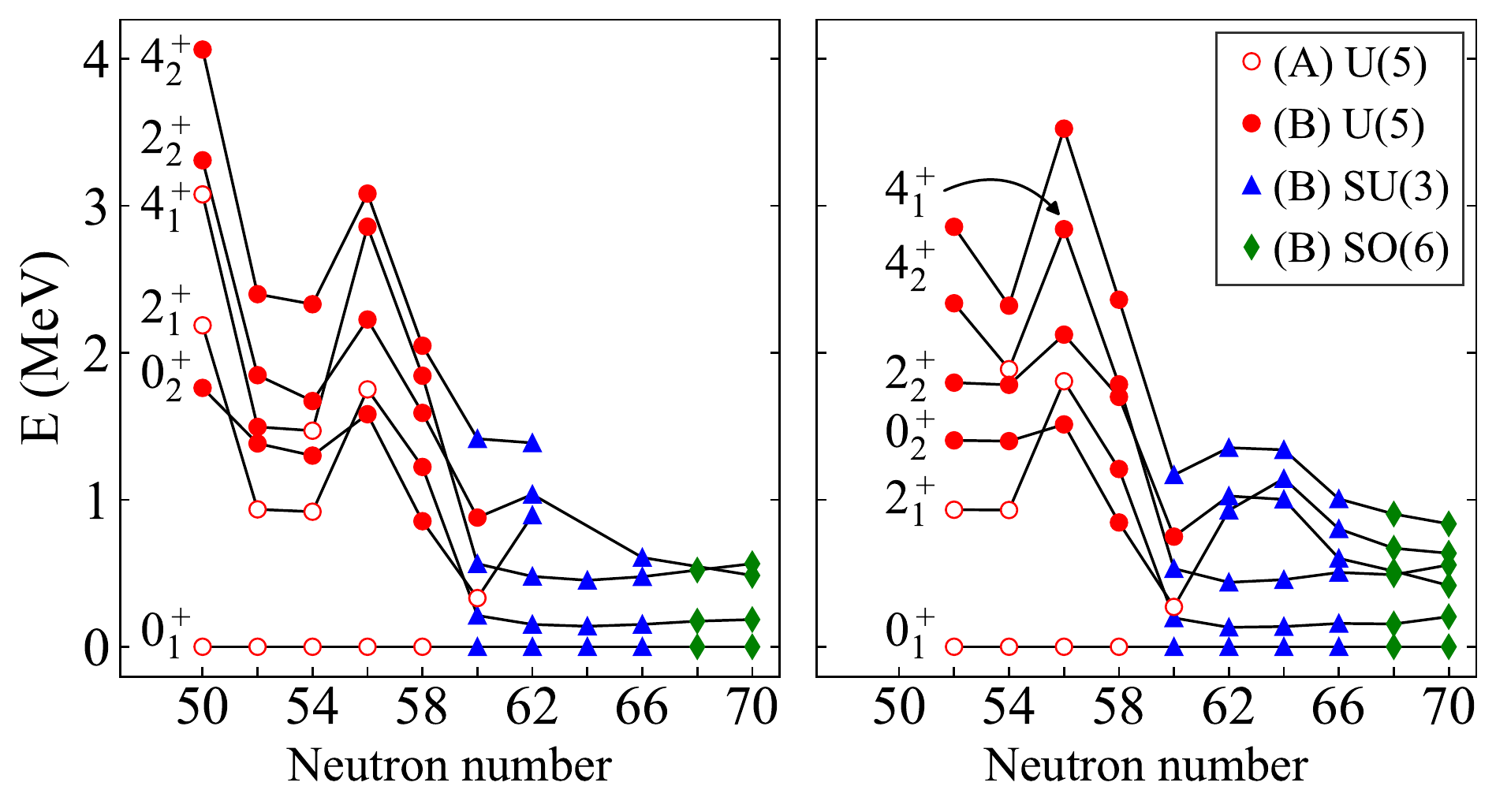}
\put (33.5,48) {(a) {\bf Exp}}
\put (56,48) {(b) {\bf Calc}}
\end{overpic}
\caption{Comparison between (a)~experimental and
(b)~calculated energy levels
$0_{1}^{+},2_{1}^{+},4_{1}^{+},0_{2}^{+},2_{2}^{+},4_{2}^{+}$.
Empty (filled) symbols indicate a state dominated by the 
normal $A$ configuration (intruder $B$ configuration), with 
assignments based on \cref{eq:decomp-cm}. 
The symbol[${\color{red}\CIRCLE}$,
${\color{blue}\blacktriangle}$,
${\color{GreenNoam}\blacklozenge}$],
indicates the closest dynamical symmetry [U(5), SU(3), 
SO(6)] to the level considered, based on 
\cref{eq:decomp-ds}. Note that the calculated values start 
at neutron number 52, while the experimental values include 
the closed shell at 50. References for the data can be 
found in 
\cite{Gavrielov2022}.\label{fig:levels}}
\end{figure}
In \cref{fig:levels}, we show a comparison between selected 
experimental and calculated levels, along with assignments 
to configurations based on \cref{eq:decomp-cm} and to the 
closest DS based on \cref{eq:decomp-ds}, for each state. In 
the 
region between neutron number 50 and 56, there appear to be 
two 
configurations, one spherical (seniority-like), ($A$), and 
one 
weakly deformed, ($B$), as evidenced by the ratio 
$R_{4/2}$, 
which is $R^{(A)}_{4/2}\cong 1.6 $ and  $R^{(B)}_{4/2} 
\cong 
2.3$ at at 52--56. 
From neutron number 58, there is a pronounced drop in 
energy for the configuration~($B$) states and at 60, the 
two configurations exchange their role, indicating a 
Type~II QPT. At this stage, the $B$ configuration appears 
to undergo a U(5)-SU(3) Type~I QPT, similarly to case of 
the Sm region~\cite{IachelloArimaBook, 
Scholten1978, Caprio2002}. 
Beyond neutron number 60, the $B$ configuration 
is strongly deformed, as evidenced by the small value of the
excitation energy of the state $2_{1}^{+}$, 
$E_{2_{1}^{+}}\!=\!139.3$~keV and by the ratio 
$R^{(B)}_{4/2}\!=\!3.24$ in $^{104}$Zr. At still 
larger neutron number 66, the ground state band becomes 
\mbox{$\gamma$-unstable} (or triaxial) as evidenced
by the close energy of the states $2_{2}^{+}$ and 
$4_{1}^{+}$, $E_{2_{2}^{+}}\!=\!607.0$~keV, 
$E_{4_{1}^{+}}\!=\!476.5$~keV, in $^{106}$Zr, 
and especially by the results $E_{4^+_1}\!=\!565$~keV and $ 
E_{2^+_2}\!=\!485$ keV for $^{110} $Zr of 
Ref.~\cite{Paul2017}, 
a signature of the SO(6) symmetry. In this region, the $B$ 
configuration undergoes a~crossover from SU(3) to SO(6).

\subsection{Evolution of configuration 
content}\label{sec:evo-conf}
\definecolor{GreenNoam}{rgb}{0,0.5,0}
\definecolor{GrayNoam}{rgb}{0.8,0.8,0.9003921568627451}
\begin{figure}[t]
\centering
\includegraphics[width=0.424\linewidth]
{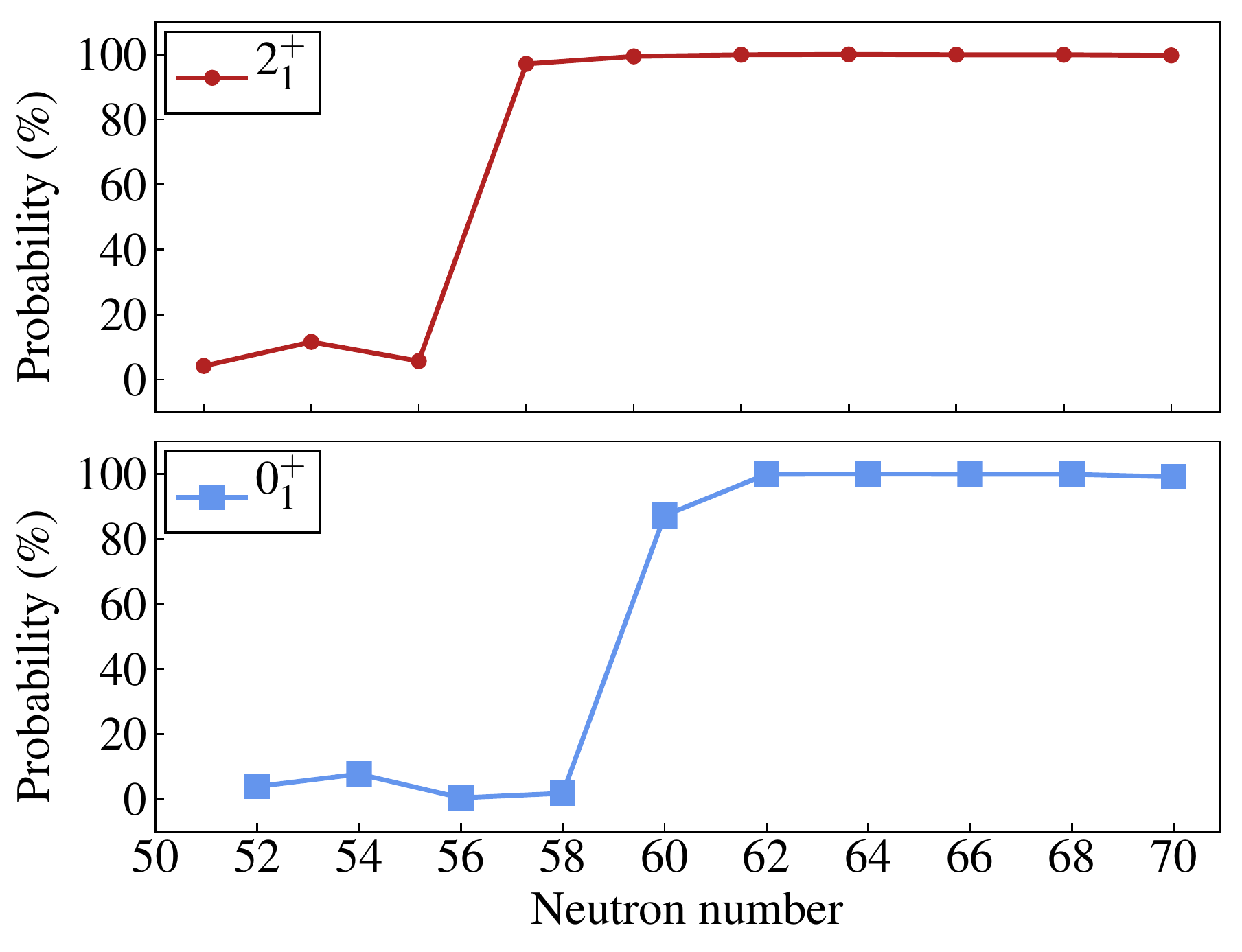}
\includegraphics[width=0.566\linewidth]
	{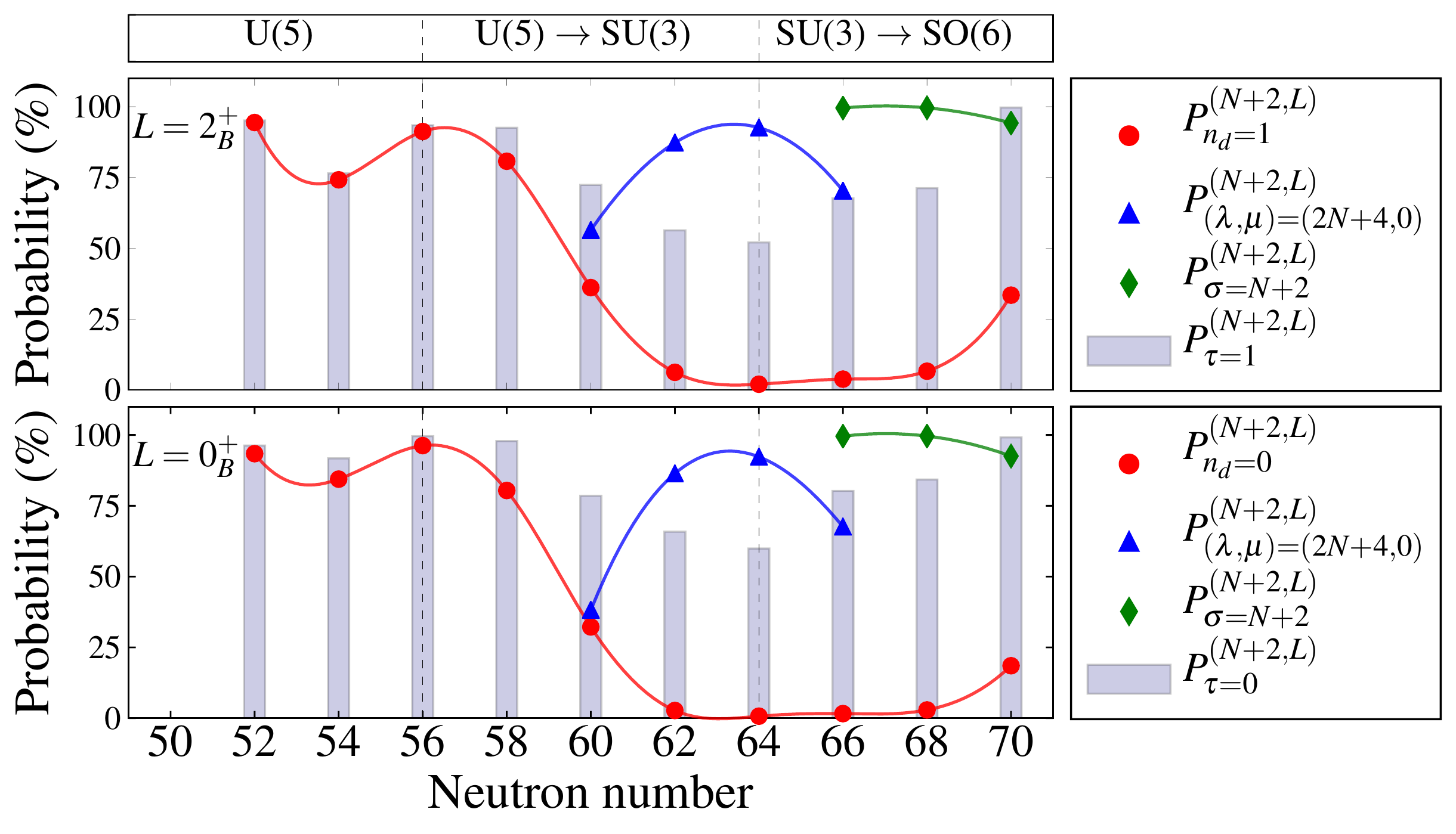}
\caption{Left panels: percentage of the wave functions 
within the intruder B-configuration [the $b^2$ probability 
in Eq.~\eqref{eq:wf}], for the ground $0^+_1$ (bottom) and 
excited $2^+_1$ (top) states in $^{92-110}$Zr.
Right panels: evolution of symmetries for the lowest $0^+$ 
(bottom) and $2^+$ (top) state of configuration $B$ along 
the Zr chain. Shown are the probabilities of selected 
components of U(5)~(${\color{red}\CIRCLE}$), 
SU(3)~(${\color{blue}\blacktriangle}$), 
SO(6)~(${\color{GreenNoam}\blacklozenge}$) and 
SO(5)~(${\color{GrayNoam}\rule[0pt]{13pt}{5pt}}$), obtained 
from \cref{eq:decomp-ds}. For neutron numbers 52--58 
(60--70), $0^+_B$ corresponds to the experimental $0^+_2$ 
($0^+_1$) state. For neutron numbers 52--56 
(58--70), $2^+_B$ corresponds to the experimental $2^+_2$ 
($2^+_1$) state.\label{fig:mixing}}
\end{figure}
We examine the configuration change for each 
isotope, by calculating the evolution of the probability 
$b^2$, \cref{eq:decomp-cm}, of the $0^+_1$ and $2^+_1$ 
states. The left panels of \cref{fig:mixing} shows the 
percentage of the wave function within the $B$ 
configuration as a function of neutron number across the Zr 
chain. 
The rapid change in structure of the $0^+_1$ state (bottom 
left panel) from the normal $A$ configuration in 
$^{92-98}$Zr (small $b^2$ probability) to the intruder $B$ 
configuration in $^{100-110}$Zr (large $b^2$ probability) 
is clearly evident, signaling a Type II QPT. The 
configuration change appears however sooner in the $2^+_1$ 
state (top left panel), which changes to configuration $B$ 
already in $^{98}$Zr, in line with \cite{Witt2018}. 
Outside 
a narrow region near neutron number 60, where the crossing 
occurs, the two configurations are weakly mixed and the 
states retain a high level of purity, especially for 
neutron number larger than 60.
\subsection{Evolution of symmetry 
content}\label{sec:evo-sym}
We examine the changes in symmetry of the lowest 
$0^+$ and $2^+$ states within the $B$ configuration, which 
undergoes a Type I QPT. In the right bottom panel of 
\cref{fig:mixing} the red dots represent the percentage of 
the U(5) $n_d\!=\!0$ component in the wave function, 
$P^{(N+2,L=0)}_{n_d=0}$ of \cref{eq:decomp-ds}. It is large 
($\approx90\%$) for neutron number 52--58 and drops 
drastically ($\approx30\%$) at 60. The drop means that 
other $n_d\!\not=\!0$ components are present in the wave 
function and therefore this state becomes deformed. 
Above neutron number 60, the $n_d\!=\!0$ component drops 
almost to zero (and rises again a little at 70), indicating 
the state is strongly deformed.
To understand the type of DS associated with the 
deformation above neutron number 60, we add in blue 
triangles the percentage of the SU(3) 
$(\lambda,\mu)=(2N+4,0)$ component, 
$P^{(N+2,L=0)}_{(\lambda,\mu)=(2N+4,0)}$ of 
\cref{eq:decomp-ds} for 60--66. For neutron number 60, it 
is moderately small ($\approx35\%$), at neutron number 62 
it jumps ($\approx85\%$) and becomes maximal at 64 
($\approx92\%$). This serves as a clear evidence for a 
U(5)-SU(3) Type~I QPT. At neutron number 66 the SU(3)
$(\lambda,\mu)\!=\!(2N+4,0)$ component
it is lowered, and one sees by the green diamonds the 
percentage of the SO(6) $\sigma=N+2$ component, 
$P^{(N+2,L=0)}_{\sigma=N+2}$ of \cref{eq:decomp-ds}. The 
latter becomes dominant for 66--70 ($\approx99\%$), 
suggesting a crossover from SU(3) to SO(6).

In order to further elaborate the Type~I QPT within 
configuration~$B$ from U(5) to SU(3) and the subsequent 
crossover to SO(6), we examine also the evolution of SO(5) 
symmetry.
The gray histograms in the right panel of \cref{fig:mixing} 
depict the probability of the $\tau\!=\!0$ component of 
SO(5), $P^{(N+2,L=0)}_{\tau=0}$ of \cref{eq:decomp-ds}, for 
$0^+_B$. 
For neutron numbers 52--56, the $0^+_B$ state is composed 
mainly of a single ($n_d\!=\!0,\tau\!=\!0$) component, 
appropriate for a with state good U(5) DS. 
For neutron number 58, the larger $\tau\!=\!0$ but smaller 
$n_d\!=\!0$ probabilities imply the presence of additional 
components with ($n_d\!\neq\!0, \tau\!=\!0$).
For neutron numbers 60--64, the $\tau\!=\!0$ probability 
decreases, implying admixtures of components with ($n_d\neq 
0, \tau\neq 0$), appropriate for a state with good SU(3) DS.
For neutron numbers 66--70, the $\tau\!=\!0$ probability 
increases towards its maximum value at 70, appropriate for 
a crossover to SO(6) structure with good SO(5) symmetry.

In the top right panel of \cref{fig:mixing} we observe a 
similar trend for the $2^+_B$ state. For neutron numbers 
52--58, it is dominated by a single 
($n_d\!=\!1,\tau\!=\!1$) component. For neutron number 60, 
$P^{(N+2,L=2^+_B)}_{n_d=1}$ is smaller than 
$P^{(N+2,L=2^+_B)}_{\tau=1}$, indicating the onset of 
deformation. For 62--64, $P^{(N+2,L=2^+_B)}_{n_d=1}$ is 
much smaller than $P^{(N+2,L=2^+_B)}_{\tau=1}$, 
implying admixtures of components with ($n_d\!\neq\!1, 
\tau\!\neq\!1$). For neutron numbers 66--70, 
$P^{(N+2,L=2^+_B)}_{n_d=1}$ remains small
but $P^{(N+2,L=2^+_B)}_{\tau=1}$ increases towards its 
maximum value at 70.

\subsection{Evolution of order parameters}
\begin{figure}[t]
\centering
\begin{overpic}[width=0.45\linewidth]{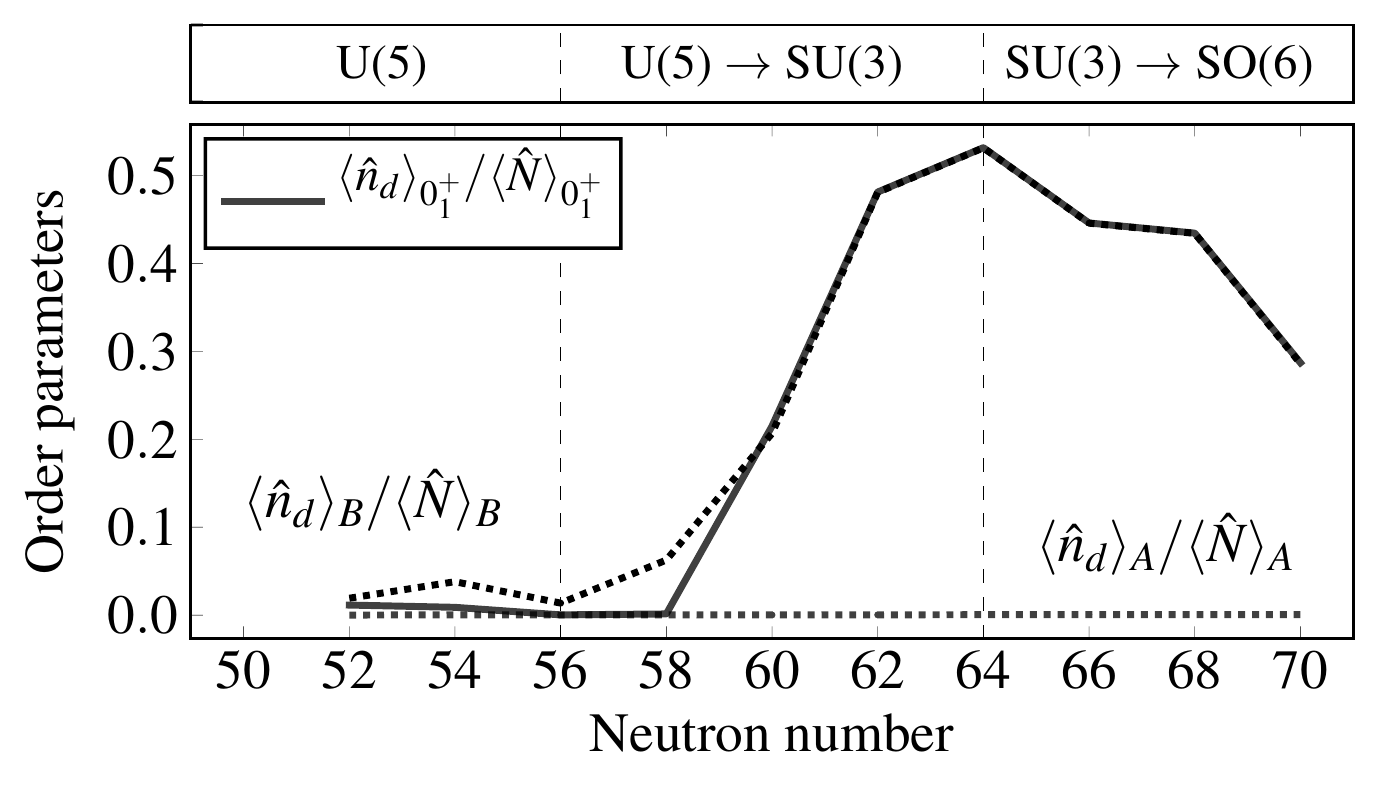}
\put (90,40) {(a)}
\end{overpic}
\begin{overpic}[width=0.45\linewidth]{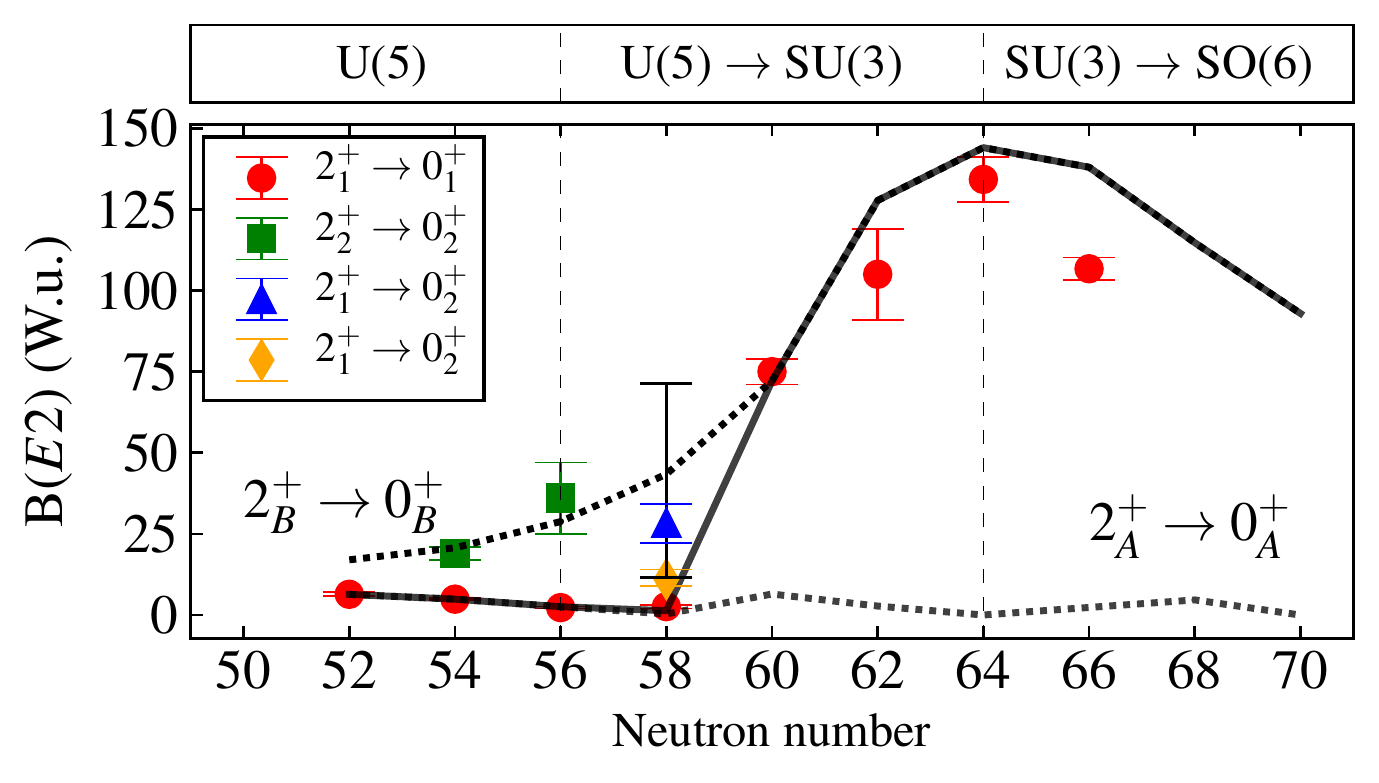}
\put (90,40) {(b)}
\end{overpic}
\caption{(a) Evolution of order parameters along the Zr 
chain, normalized (see text).
(b) $B(E2)$ values in W.u. for $2^+\rightarrow0^+$ 
transitions in the Zr chain. The solid line (symbols 
${\color{red}\CIRCLE},~{\color{GreenNoam}\blacksquare},
~{\color{blue}\blacktriangle},
~{\color{orange}\blacklozenge}$)
 denote calculated results (experimental results). Dotted 
lines denote calculated $E2$ transitions within a 
configuration.
The data for $^{94}$Zr, $^{96}$Zr, $^{100}$Zr, $^{102}$Zr 
and ($^{104}$Zr, $^{106}$Zr) are taken from 
\cite{Chakraborty2013}, \cite{Kremer2016}, 
\cite{Ansari2017}, \cite{NDS.110.1745.2009}, 
\cite{Browne2015b}, respectively.
For $^{98}$Zr (neutron number 58), the experimental values 
are from \cite{Karayonchev2020} 
({\color{orange}$\blacklozenge$}), from \cite{Singh2018} 
({\color{blue}$\blacktriangle$}), and the upper and lower 
limits (black bars) are from 
\cite{Ansari2017, Witt2018}.\label{fig:nd-be2}}
\end{figure}

The configuration and symmetry analysis of 
\cref{sec:evo-conf,sec:evo-sym} suggest a situation of 
simultaneous occurrence of Type~I and Type~II QPTs. 
The order parameters can give further insight to these QPTs.
Fig.~\ref{fig:nd-be2}(a) shows the evolution along 
the Zr chain of the order parameters 
($\braket{\hat{n}_d}_{A},\,\braket{\hat{n}_d}_{B}$ in 
dotted and $\braket{\hat{n}_d}_{0^{+}_1}$ in solid lines),
normalized by the respective boson numbers,
$\braket{\hat{N}}_A\!=\!N$, 
$\braket{\hat{N}}_B\!=\!N\!+\!2$,
$\braket{\hat{N}}_{0^{+}_1}\!=\!a^2N\!+\!b^2(N\!+\!2)$.
The order parameter $\braket{\hat{n}_d}_{0^{+}_1}$ is close 
to $\braket{\hat{n}_d}_A$ for neutron number 52--58 and 
coincides with $\braket{\hat{n}_d}_B$ at 60 and above. The 
clear jump and change in configuration content from 58 to 
60 indicates a Type~II phase transition~\cite{Frank2006}, 
with weak mixing between the configurations.
Configuration~$A$ is spherical for all neutron numbers, and 
configuration~$B$ is weakly-deformed for neutron number 
52--58. From neutron number 58 to 60 we see a sudden 
increase in $\braket{\hat{n}_d}_B$ that continues towards 
64, indicating a U(5)-SU(3) Type~I phase transition. Then, 
we observe a decrease from neutron number 
66 onward, due in part to the crossover from SU(3) to SO(6) 
and in part to the shift from boson particles to boson 
holes after the middle of the major shell 50--82. 
These conclusions are stressed by an analysis of other 
observables~\cite{Gavrielov2022}, in particular, the 
$B(E2)$ values. As shown in Fig.~\ref{fig:nd-be2}(b), the 
calculated $B(E2)$'s agree with the experimental values and 
follow the same trends as the respective order parameters.

\subsection{Classical analysis}\label{sec:classical}
\begin{figure}[]
\centering
\begin{overpic}[width=0.18\linewidth]{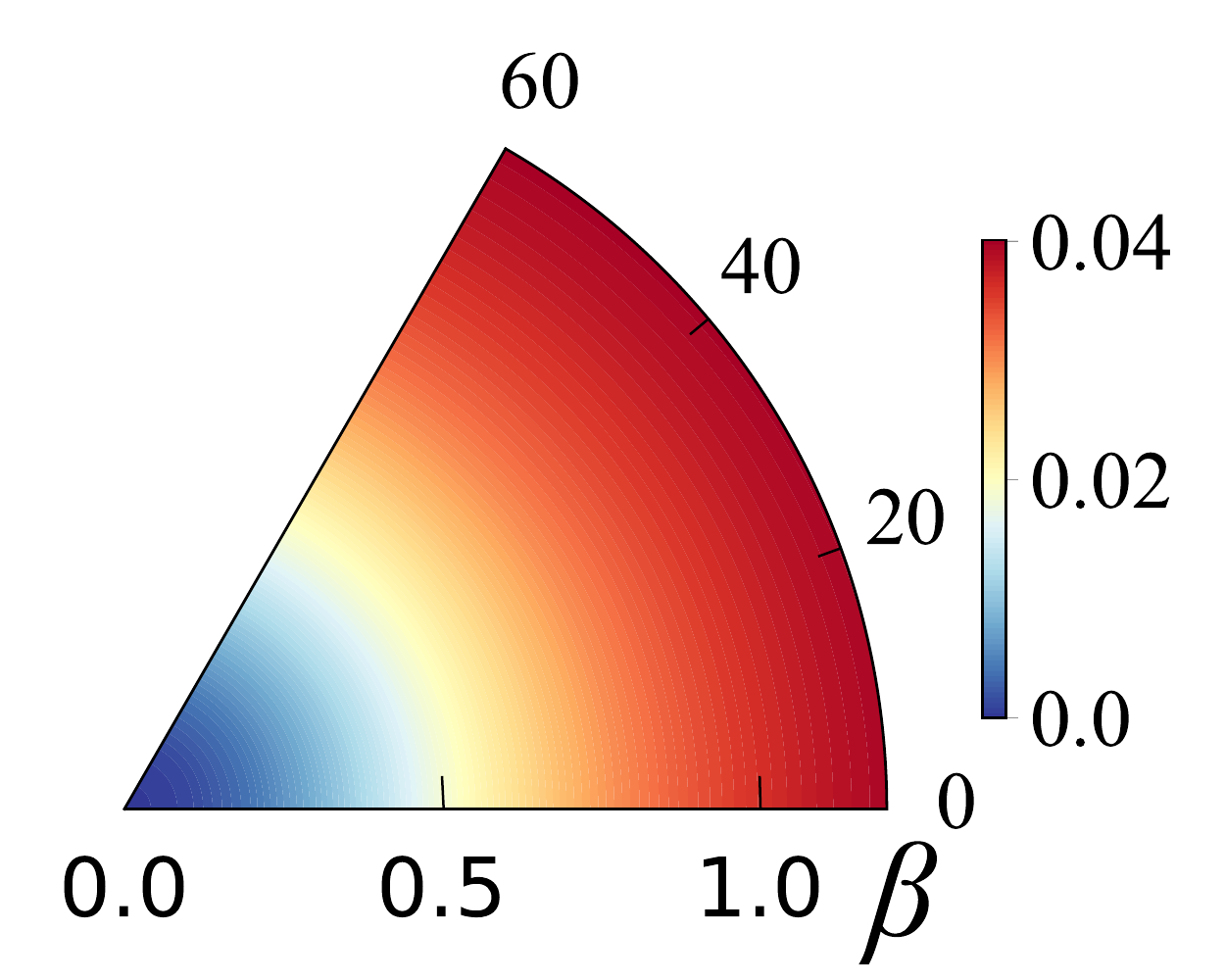}
\put (-5,69) {(a)}
\put (-5,50) { $^{92}$Zr}
\put (60,70) {\small$\gamma$(deg)}
\end{overpic}
\begin{overpic}[width=0.18\linewidth]{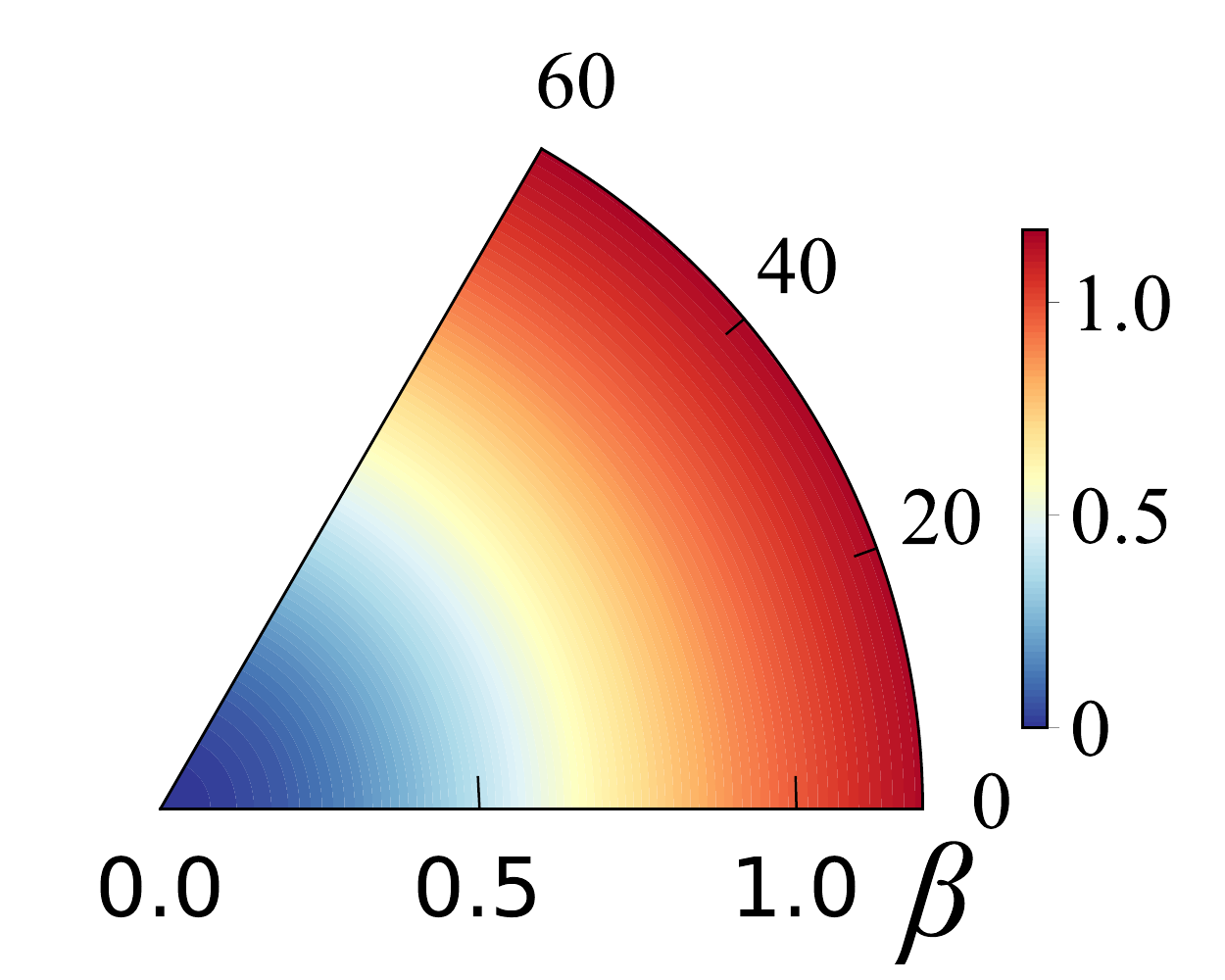}
\put (-5,69) {(b)}
\put (-5,50) { $^{94}$Zr}
\put (60,70) {\small$\gamma$(deg)}
\end{overpic}
\begin{overpic}[width=0.18\linewidth]{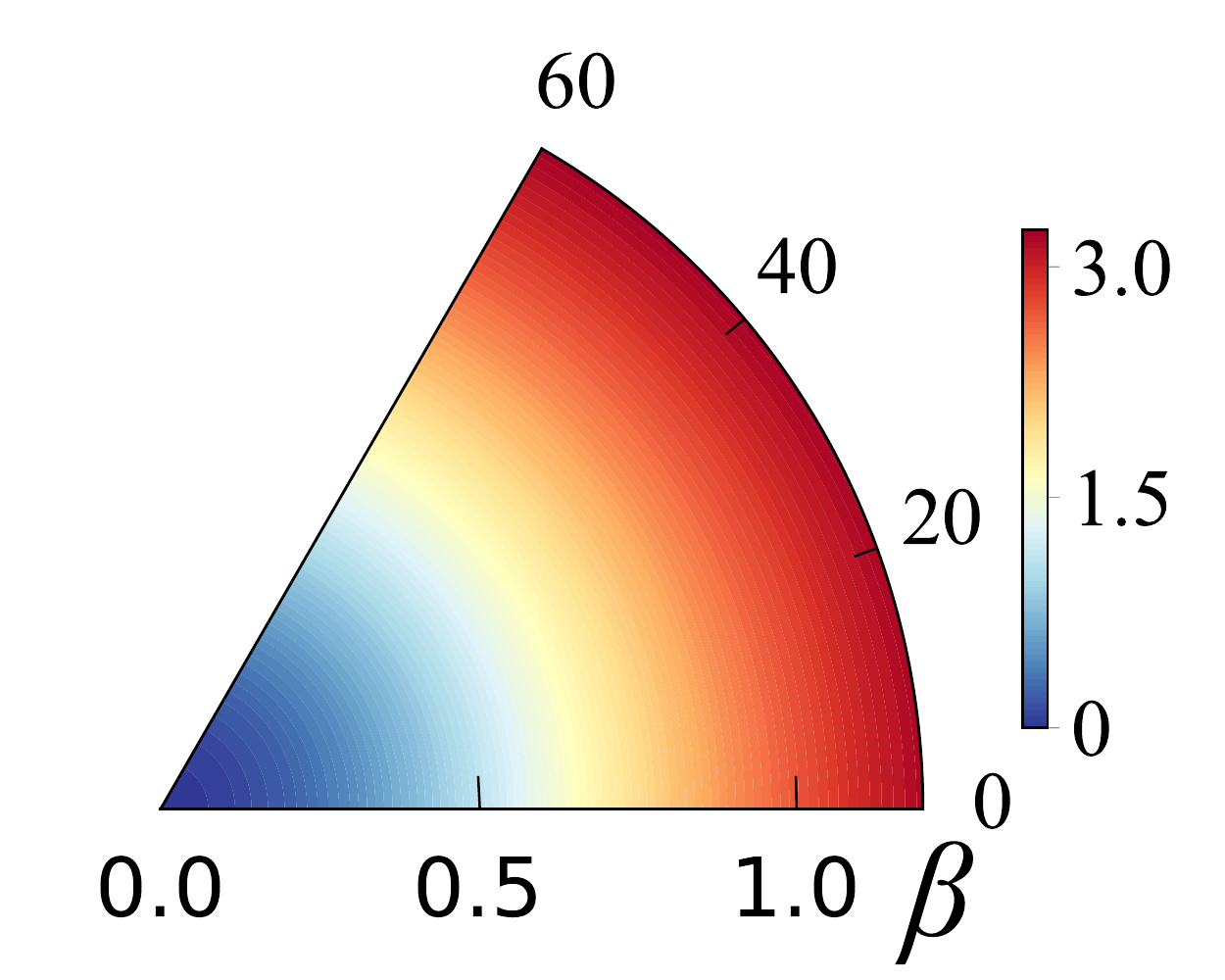}
\put (-5,69) {(c)}
\put (-5,50) { $^{96}$Zr}
\put (60,70) {\small$\gamma$(deg)}
\end{overpic}
\begin{overpic}[width=0.18\linewidth]{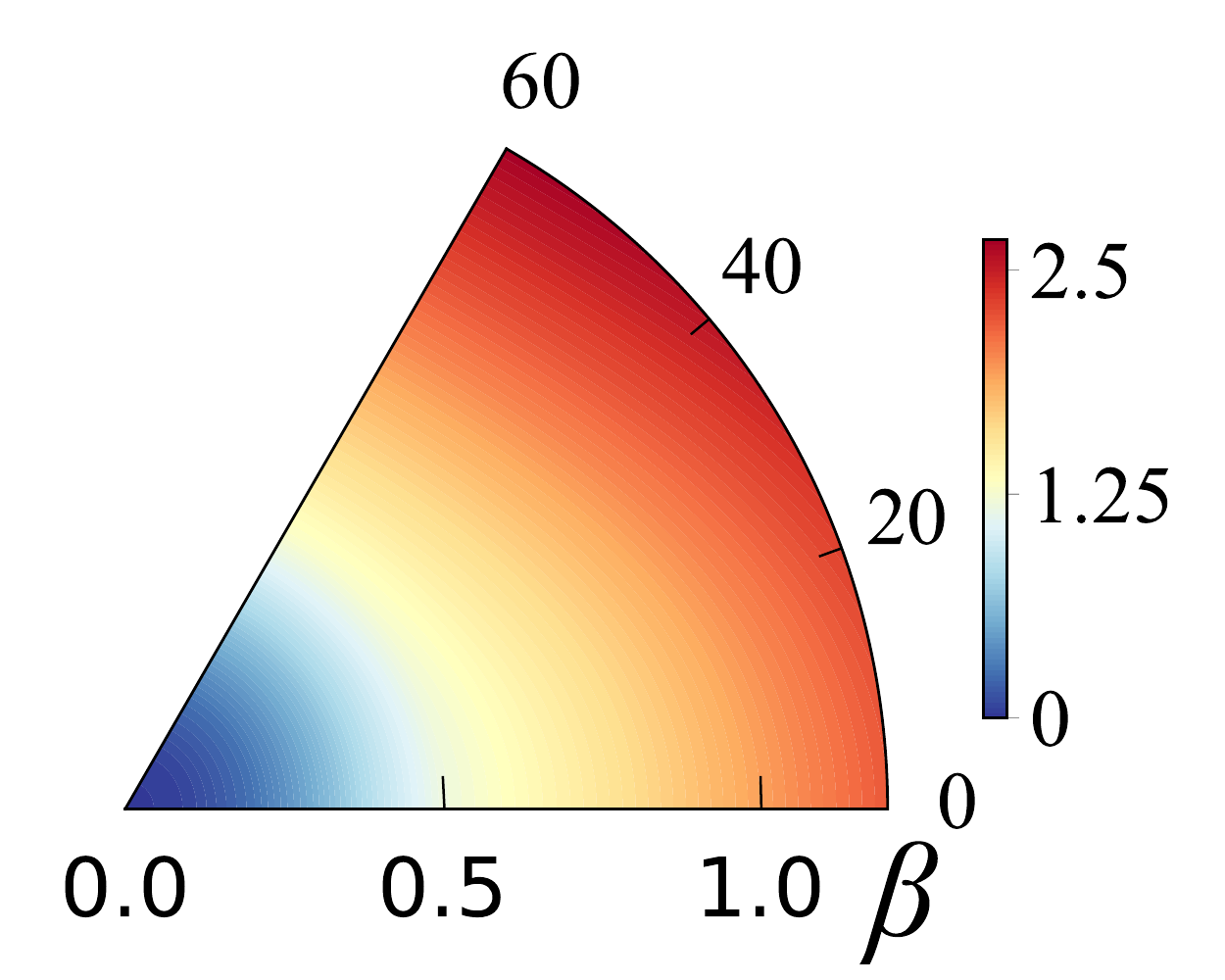}
\put (-5,69) {(d)}
\put (-5,50) { $^{98}$Zr}
\put (60,70) {\small$\gamma$(deg)}
\end{overpic}
\begin{overpic}[width=0.18\linewidth]{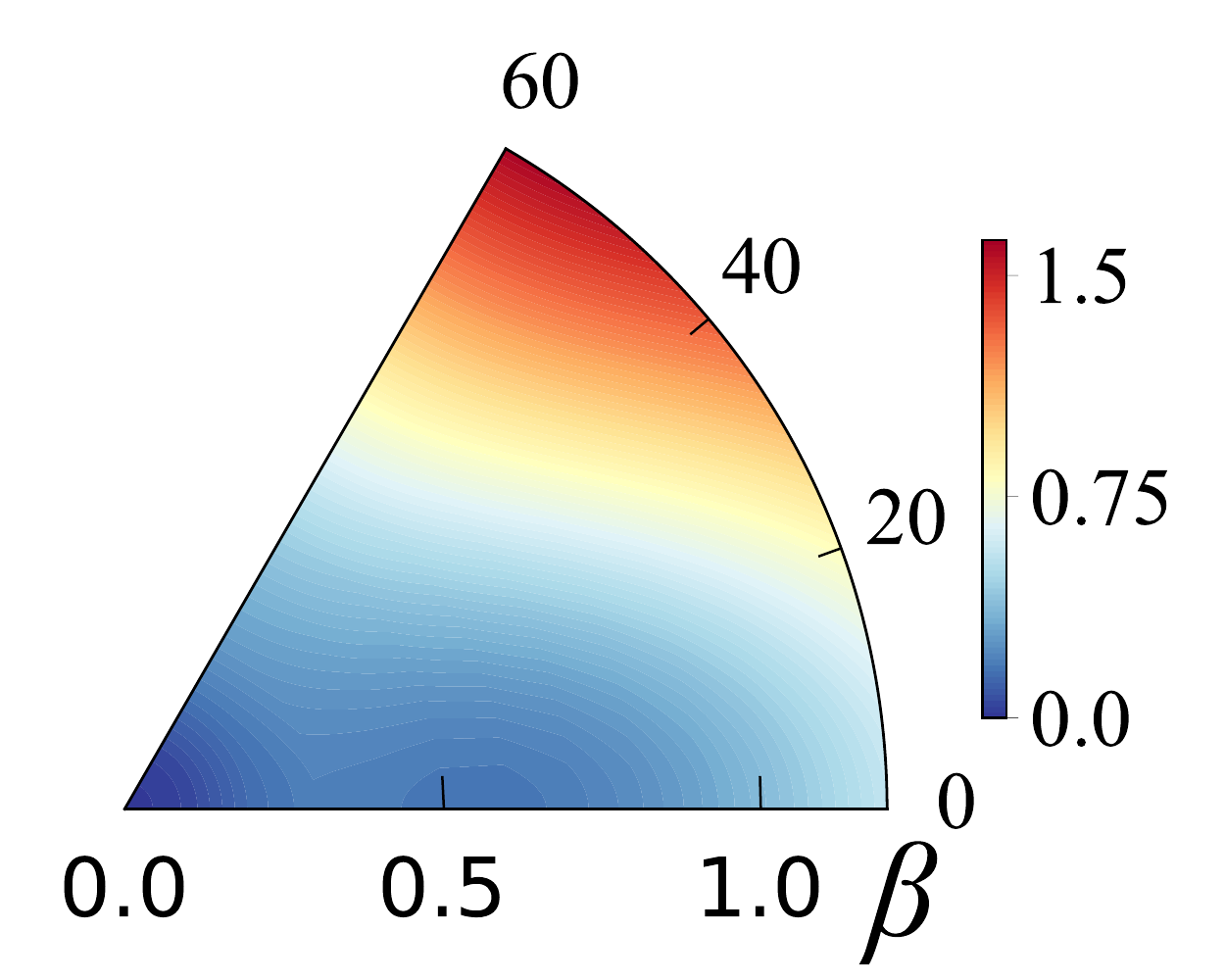}
\put (-5,69) {(e)}
\put (-5,50) { $^{100}$Zr}
\put (60,70) {\small$\gamma$(deg)}
\end{overpic} \\ 
\begin{overpic}[width=0.18\linewidth]{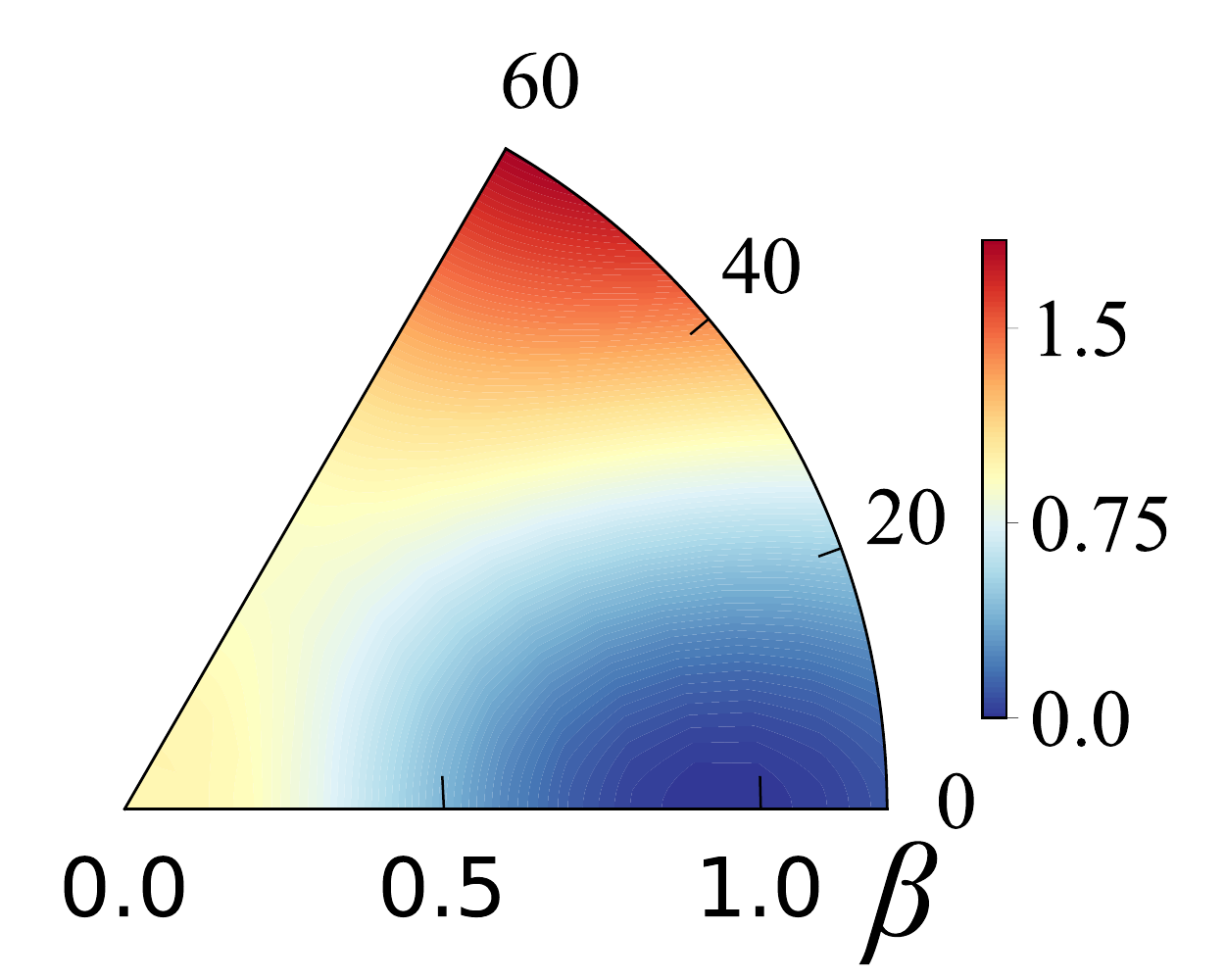}
\put (-5,69) {(f)}
\put (-5,50) { $^{102}$Zr}
\put (60,70) {\small$\gamma$(deg)}
\end{overpic}
\begin{overpic}[width=0.18\linewidth]{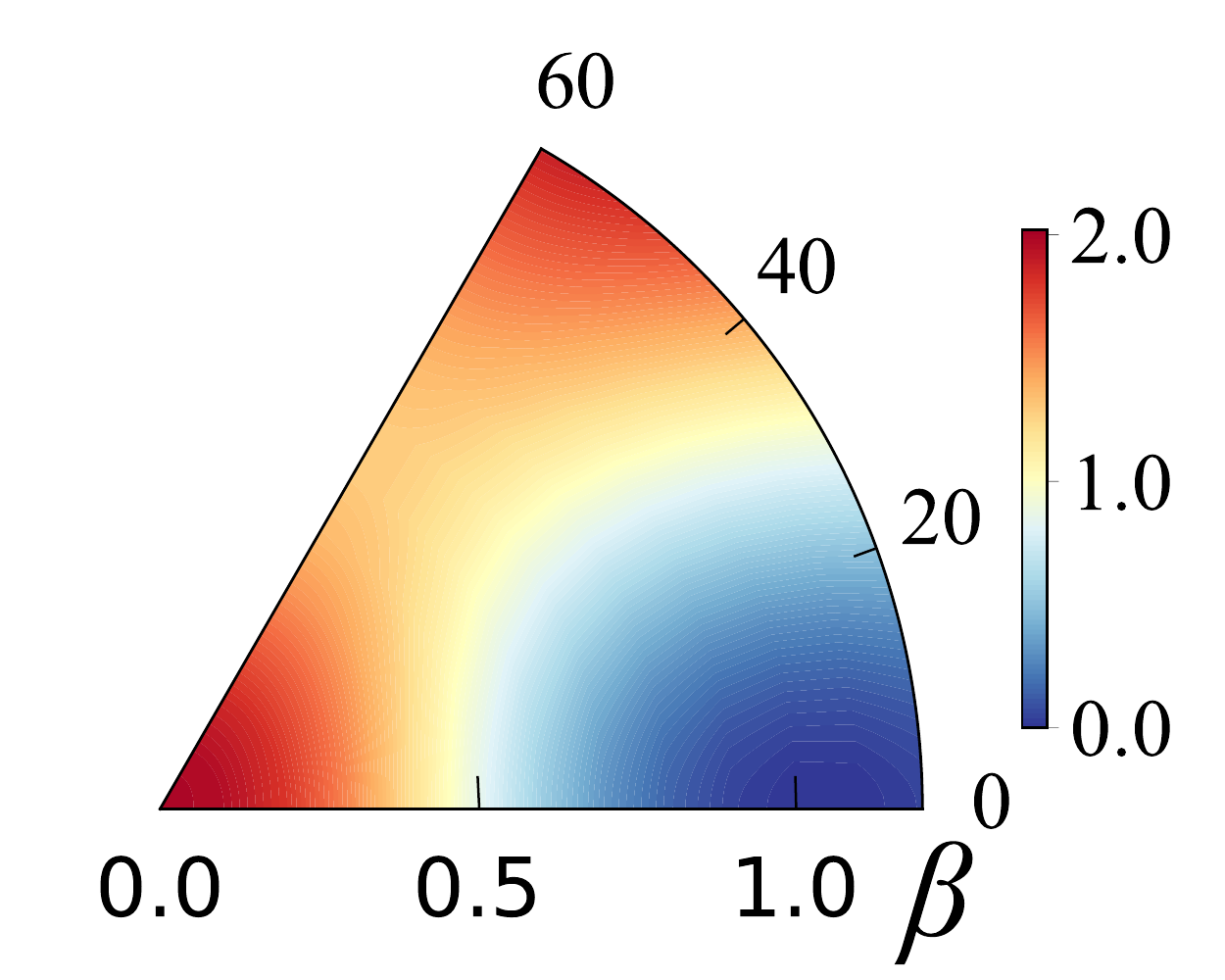}
\put (-5,69) {(g)}
\put (-5,50) { $^{104}$Zr}
\put (60,70) {\small$\gamma$(deg)}
\end{overpic}
\begin{overpic}[width=0.18\linewidth]{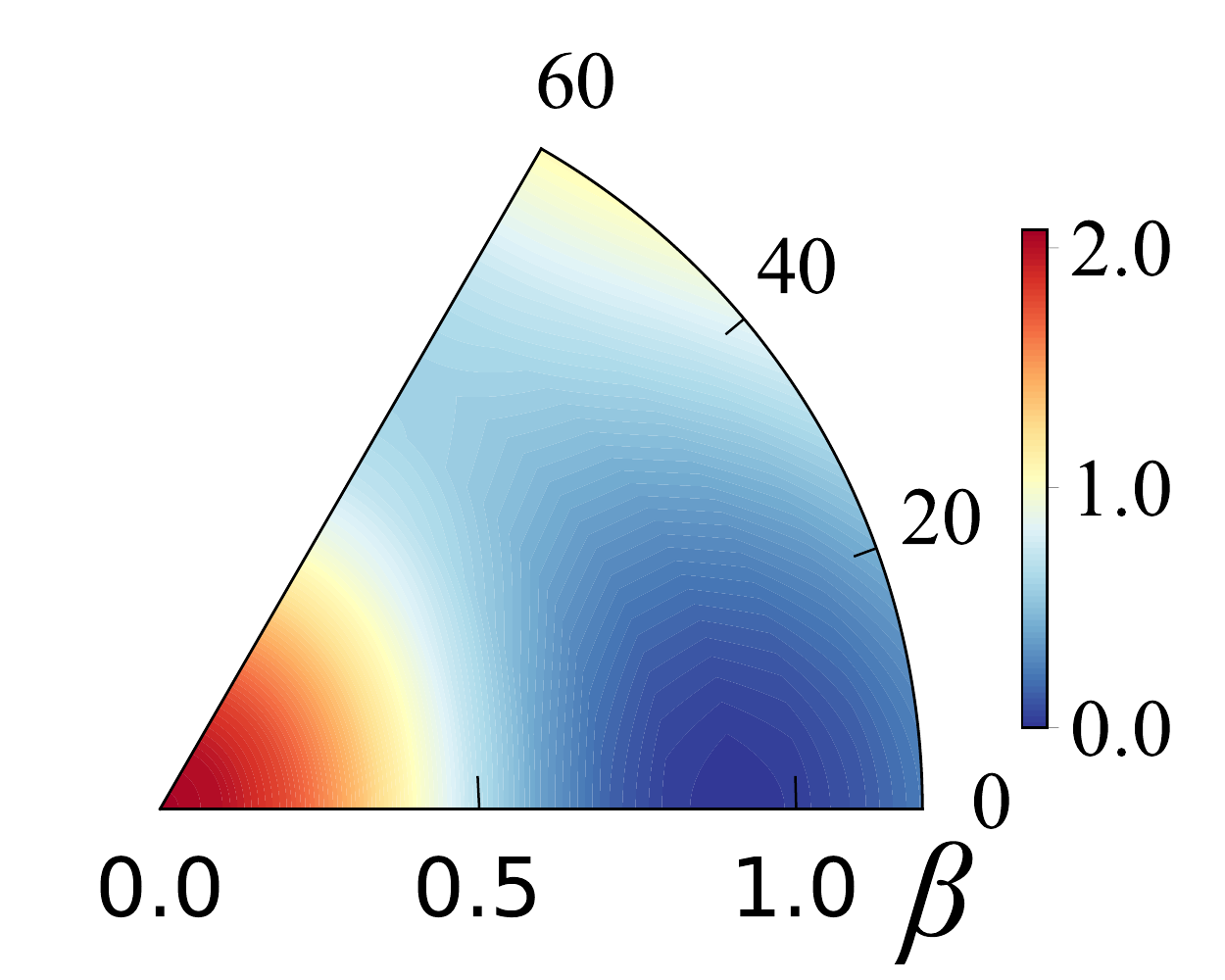}
\put (-5,69) {(h)}
\put (-5,50) {$^{106}$Zr}
\put (60,70) {\small$\gamma$(deg)}
\end{overpic}
\begin{overpic}[width=0.18\linewidth]{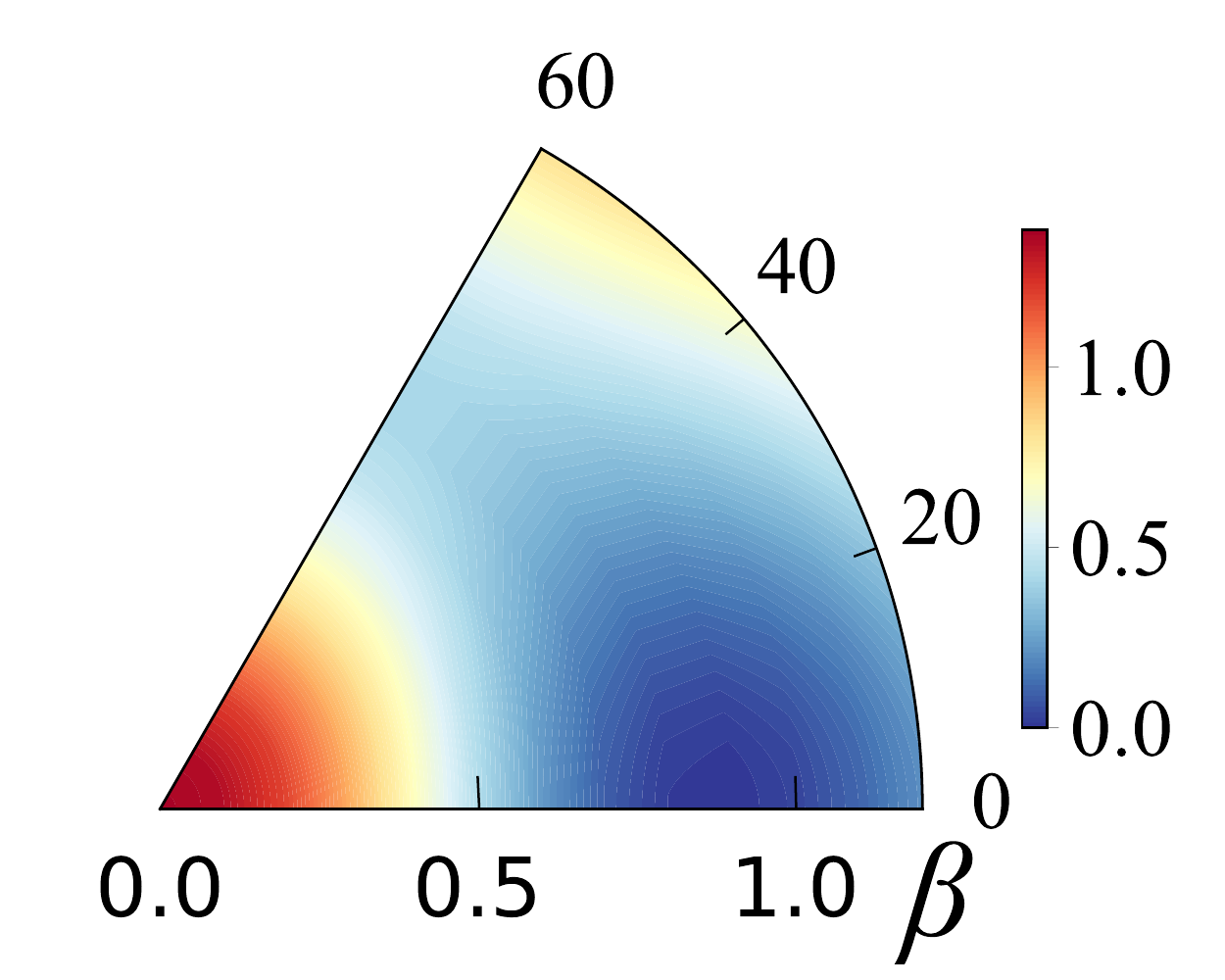}
\put (-5,69) {(i)}
\put (-5,50) { $^{108}$Zr}
\put (60,70) {\small$\gamma$(deg)}
\end{overpic}
\begin{overpic}[width=0.18\linewidth]{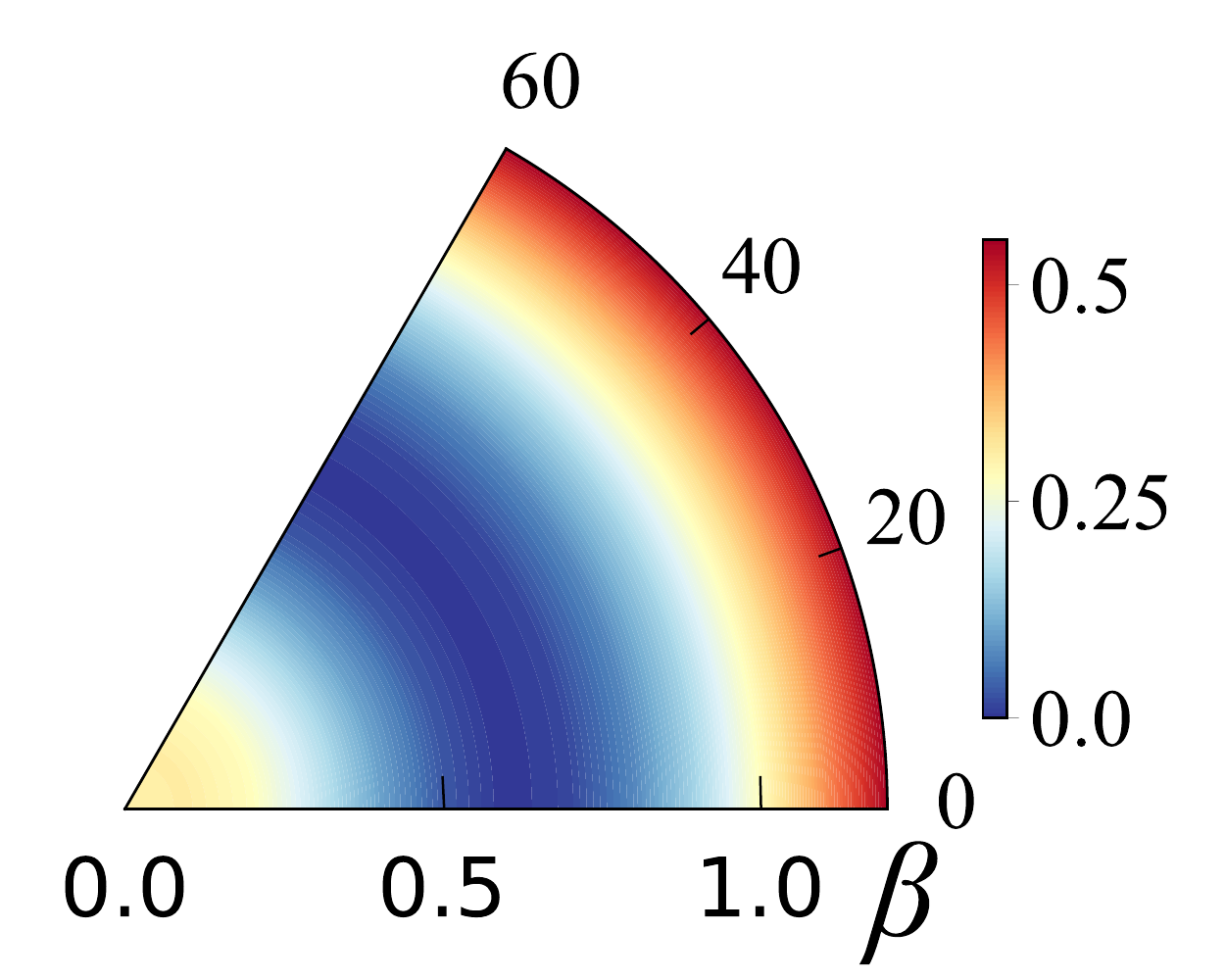}
\put (-5,69) {(j)}
\put (-5,50) { $^{110}$Zr}
\put (60,70) {\small$\gamma$(deg)}
\end{overpic}
\caption{\label{Eminus}
\small
Contour plots in the $(\beta ,\gamma)$ plane of the lowest 
eigen-potential surface, $E_{-}(\beta ,\gamma)$, for the 
$^{92-110}$Zr isotopes.}
\end{figure}
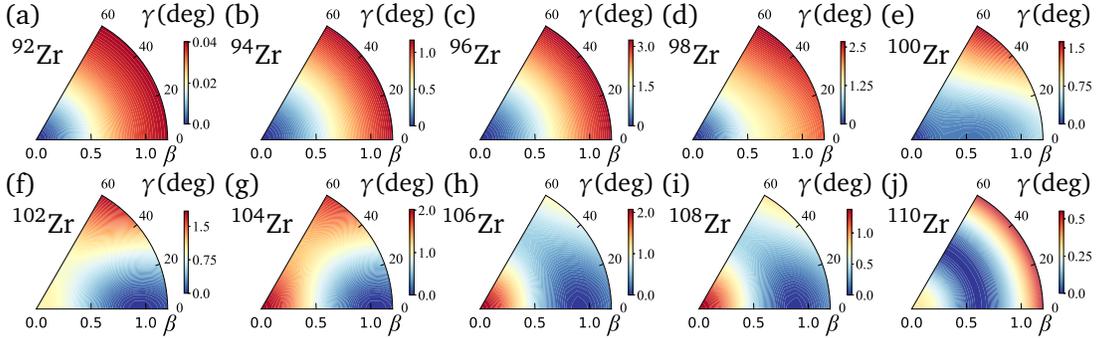
In Fig.~\ref{Eminus}, we show the calculated 
lowest eigen-potential $E_{-}(\beta ,\gamma )$, which is 
the lowest eigenvalue of the matrix \cref{eq:surface-mat}. 
These classical potentials confirm the quantum results, as 
they show a transition from spherical ($^{92-98}$Zr), 
Figs.~\ref{Eminus}(a)-(d), to a double-minima potential 
that is almost flat-bottomed at $^{100}$Zr, 
Fig.~\ref{Eminus}(e), to prolate axially deformed 
($^{102-104}$Zr), Figs.~\ref{Eminus}(f)-(g), and finally to 
$\gamma $-unstable ($^{106-110}$Zr), 
Figs.~\ref{Eminus}(h)-(j).

\section{Conclusions and outlook}
The algebraic framework of the IBM-CM allows us to examine 
QPTs using both quantum and classical analyses.
We have employed this analysis to the Zr isotopes with 
$A\!=\!92\text{--}110$, which exhibit a complex structure 
that 
involves a shape-phase transition within the intruder 
configuration (Type~I QPT) and a configuration-change 
between 
normal and intruder (Type~II QPT), namely IQPTs. This was 
done 
by analyzing the energies, configuration and symmetry 
content of 
the wave functions, order parameters and $E2$ transition 
rates, 
and the energy surfaces. 
Further analysis of other observables supporting this 
scenario 
is presented in \cite{Gavrielov2022}. Recently, we have 
also 
exemplified the notion IQPTs in the odd-mass $_{41}$Nb 
isotopes 
\cite{Gavrielov2022c} and it would be interesting to 
examine the 
notion of IQPTs in other even-even and odd-mass chains of 
isotopes in the $Z\approx40,~A\approx100$ region and other 
physical systems.

\section*{Acknowledgments}
This work was done in collaboration with 
F.~Iachello (Yale university) and A.~Leviatan (Hebrew 
university).


\paragraph{Funding information}
This work was supported in part by the US-Israel Binational 
Science Foundation Grant No. 2016032 and the Israel Academy 
of Sciences for a Postdoctoral Fellowship Program in 
Nuclear Physics.
\bibliography{algebraic_iqpt.bib}
\end{document}